\documentclass[review]{elsarticle}

\usepackage{lineno,hyperref}

\usepackage{lineno,hyperref}
\usepackage{amsmath}
\usepackage{graphicx}
\usepackage{dsfont}
\usepackage{amssymb}
\usepackage{subfigure}
\usepackage{setspace}
\usepackage{caption}
\usepackage{xcolor}
\usepackage{soul}
\newcommand{\volume}{\mathop{\ooalign{\hfil$V$\hfil\cr\kern0.08em--\hfil\cr}}\nolimits}

\definecolor{darkblue}{rgb}{0.2,0.2,1.0}

\modulolinenumbers[5]

\journal{Journal of Ocean Engineering}

\begin{document}

\begin{frontmatter}

\title{DRAFT: Wave Excitation Force Estimation of Wave Energy Floats Using Extended Kalman Filters}
%\tnotetext[mytitlenote]{Fully documented templates are available in the elsarticle package on \href{http://www.ctan.org/tex-archive/macros/latex/contrib/elsarticle}{CTAN}.}

%% Group authors per affiliation:

%% Group authors per affiliation:
\author{Andrew F. Davis\fnref{myfootnote,ourfootnote}}
%\address{Radarweg 29, Amsterdam}
\fntext[myfootnote]{Corresponding author: AndrewDavis@letu.edu Department of Mechanical Engineering LeTourneau University, Longview TX, USA}
\fntext[ourfootnote]{Department of Mechanical Engineering University of Washington, Seattle WA, USA}

\author{Brian C. Fabien\fnref{ourfootnote}}

\begin{abstract}
In many advanced control strategies the wave excitation force is key to determining the control input. However, it is often difficult to measure the excitation force on a Wave Energy Converter (WEC). The use of Kalman filters to estimate the wave excitation force based on readily available measurement data can potentially fill the gap between the development of WEC control strategies and the data that is available. Two different estimation methods using an nonlinear Extended Kalman Filter are tested on experimental wave tank data for a heaving semi-submerged float. The first method relies on directly including the excitation force as a state in the first order dynamics---which allows the ``random walk'' of the Kalman filter to identify an estimate of the excitation force. The second method of estimation involves modeling the wave excitation force as a harmonic oscillator comprised of sinusoidal components. Both methods are evaluated for a variety of incident waves and additional sensitivity analyses are performed to investigate the susceptibility of these estimation methods to changes in the model, measurement noise, and sampling rate.
\end{abstract}


\end{frontmatter}

%\linenumbers

\section{Introduction} \label{sec:introl}
Wave Energy Converters (WECs) are devices that generate electricity from ocean waves. Some non-exaustive reviews of the widely varying devices and modeling methods for WECs can be found in \cite{Day2015,Falnes2007,Li2012,Folley2016}.\\
\indent This work will focus specifically on point absorbing wave energy converters. Point absorbing type WECs rely on a small body, when compared to the incident wavelength, to be actuated primarily in a single degree of freedom to achieve power generation. A point absorbing WEC was chosen for this work as point absorbers typically offer significant deployment versatility and simplified heave dynamics.\\
\indent Modern control strategies rely on state information to calculate an optimal control input. As addressed in \cite{Brekken2011} and \cite{Ling2015}, this is possible through the use of Kalman filters as optimal state observers. Kalman filters can also be used to estimate the wave excitation force which is necessary for many controllers \cite{Ling2015,Nguyen2017}. Even though Kalman filters are commonly used to estimate the states for control systems, the implementation of a Kalman filter on a WEC is not trivial. This work addresses some of the current barriers to state observation and excitation force estimation of wave energy converters using an Extended Kalman Filter (EKF), which is the nonlinear adaptation of the Kalman filter. \\
\indent The use of an EKF to determine wave excitation force using external pressure sensors is shown in \cite{Abdelkhalik2016}. However, the implementation of additional external pressure sensors may be impractical for many commercial WEC applications. Alternatively, \cite{Ling2015} shows a way to use a Kalman filter to estimate the exitation force by modeling the excitation force as a disturbance in the WEC equations of motion. The disturbance used in \cite{Ling2015} describes the excitation force as a sum of sinusoids as is typical in WEC modeling \cite{Folley2016}. This method of estimating the wave excitation force will be referred to as the \textit{disturbance estimation method}. Alternatively, a linear Kalman filter is implemented in \cite{Nguyen2017} where the excitation force is identified by including the force as a state with no dynamics and thereby allowing the ``random walk'' of the Kalman filter to identify the best estimate of the state directly. This methodology of implementing the Kalman filter to identify the wave excitation force will be referred to as the \textit{direct estimation method}. The work of \cite{Nguyen2017} also develops a receding horizon estimator which requires the solution of a dynamic programming problem at each time step. Estimation using the receding horizon method results in increased computational complexity but provides very little phase lag between the estimated and true excitation forces \cite{Nguyen2017}. A similar disturbance model estimation method to \cite{Ling2015} is \cite{Ringwood2017}, where a linear Kalman filter is used on simulation data to perform an investigation into excitation force estimation and prediction. The direct estimation method shown in \cite{Nguyen2017} applies the same principle that is used for drag estimation in \cite{Davis2018} and unknown parameter identification in \cite{Crassidis2011}. The work of \cite{Faedo2017} provides an overview of control algorithms, many of which rely on the use of a linear Kalman filter. Excitation force prediction is of primary interest in the overview \cite{Faedo2017}. 

\indent This work describes the implementation of an extended Kalman filter that uses a nonlinear continuous time model and discrete time measurements. The use of a nonlinear Kalman filter enables the inclusion of the nonlinear Morison drag term, that has become a commonly included in many lumped parameter models and WEC modeling software packages \cite{Bhinder2015,Combourieu2015}. The implementation and sensitivity analysis of wave excitation force estimation with an extended Kalman filter on experimental data from \cite{Bosma2015} is described using both the direct estimation and disturbance estimation methods. The considerations that need to be taken into account when implementing these two methods with a nonlinear model are given, and the sensitivity of the estimation methods to various conditions are investigated.

\indent The work described in this paper is differentiated from previous literature in this field by providing the following contributions:
\begin{itemize}
\item The excitation force estimation is performed on experimental wave tank data.
\item The implementation of a nonlinear Extended Kalman Filter made use of two different methods of adapting the Kalman filter to estimate the excitation force.
\item The inclusion of an estimated velocity for the water surrounding the float allows for the nonlinear Morison drag term to be used in the model. 
\item The sensitivity of the estimation algorithm to measurement noise, wave climate, sampling rate, radiation realization, and the number of frequencies that are used to model the excitation force are studied.
\item  An equal energy approach is taken to choose the wave frequencies that comprise the disturbance model of the wave excitation force.
\end{itemize}

This work provides insight into the implementation of a nonlinear Kalman filter with experimental data for wave energy conversion. Reproducible methods are described to facilitate implementation of the EKF for excitation force estimation and an investigation of the limitations of the EKF is performed to help ensure the suitability of the application. The purpose of this work is to enhance the performance of state observers when estimating the wave excitation force acting on WECs with the goal of enabling more effective feedback control.

Advanced control strategies such as latching control or Model Predictive Control (MPC) offer increased power production when compared to traditional WEC control strategies \cite{Hong2014,Brekken2011}. Optimal power absorption is achieved when the phase of the WEC is properly aligned with the wave excitation force, to this end latching control uses an actuator to lock a WEC in a specific configuration long enough for the WEC to achieve the optimal phase with the excitation force \cite{Falnes2007}. Model predictive control and Nonlinear Model Predictive Control (NMPC) are methods that solve a series of successive, small time horizon, optimal control problems \cite{Tom2015}. The optimal control problem is solved for a future timespan and takes into account the current wave climate when determining the optimal control input. However, a short prediction of incident wave elevation or wave excitation force is necessary in order to implement this type of controller \cite{Fusco2012}. As the wave excitation force estimation is enhanced forecasting methods can be better implemented to provide the necessary time horizons for predictive and latching control methods.

The dynamic model used in this work is described in section \ref{sec:wecModel}. 
The experimental tests used in the development and implementation of the estimation algorithms is described in section \ref{sec:experiment}. A description of the identification algorithms used is given in section \ref{sec:algorithm}. Excitation force estimation results and sensitivity analyses are given in section \ref{sec:results}. Finally, concluding remarks are presented in section \ref{sec:conclusion}.

\section{Wave Energy Converter Model} \label{sec:wecModel}
\indent The potential based time domain integro-differential equation analyzed in \cite{Babarit2012} has become the most prolific lumped parameter model for a WEC Equation of Motion (EOM). However, semi-emperical models or simplified lumped parameter models are also possible for WECs as shown in the detailed modeling description given in \cite{Davis2018}. The equation of motion of a semi-submerged body in heave is given by
\begin{align}
M_v\ddot{y}&=F_B+F_D+F_R+F_M+F_{ex}. \label{eq:simpleEOM}
\end{align}
where the inertial force of the object is the product of the ``virtual mass", $M_v$, and the heave acceleration, $\ddot{y}$. This ``virtual mass" is the sum of the mass of the body and the infinite frequency added mass of the entrained fluid, as is traditionally used in the Cummins formulation \cite{Cummins1962}. The terms used in (\ref{eq:simpleEOM}) are defined as: $F_B$ is the hyrdostatic force on the object, $F_D$ is the viscous drag force, $F_R$ is the radiation damping, $F_M$ is the restoring force from mooring lines, $F_{ex}$ is the wave excitation force, and the displacement of the body is $y$. The dot notation represents derivatives with respect to time, for example, $\dot{y}=\frac{dy}{dt}$ and $\ddot{y}=\frac{d^2y}{dt^2}$. Discussion of the simplification of a general equation of motion to the form shown in (\ref{eq:simpleEOM}) can be found in \cite{Davis2018}.\\
\indent The potential based force terms of (\ref{eq:simpleEOM}) are developed in agreement with the Cummins equation \cite{Cummins1962} which was originally developed to describe the motion of ships and is now applied to semi-submerged bodies. The viscous drag term is the lone nonlinear term in (\ref{eq:simpleEOM}) as described in \cite{Babarit2012,Bhinder2015} and the inertial, excitation, restoring, and radiation forces are assumed to be linear for the following development. 
\subsection{Mooring and Hydrostatic Forces $F_M$ and $F_B$}
\indent The forces resulting from mooring and umbilical cables are modeled together as the product of a constant parameter and the body displacement as given by the linear relationship
\begin{align}
F_M=-K_{mu} y, \label{eq:F_M}
\end{align}
where $K_{mu}$ is lumped linear spring constant describing the forces acting on the body by the mooring lines and umbilical cable and $y$ is the displacement from the static equilibrium point. By modeling the response of the body from the equilibrium position, as with the hydrostatic restoring force, only a linear relationship is needed to determine the vertical force contribution of the cables as a function of heave position. The constant can be determined experimentally using load cells and displacement measurements, described in section \ref{sec:experiment}\\
\indent The hydrostatic restoring force is determined by 
\begin{align}
F_B=-\rho gAy, \label{eq:F_B}
\end{align}
where $A$ is the projected area that is being submerged, $\rho$ is the density,$g$ is the gravitational acceleration constant, and $y$ is the displacement from the static equilibrium. It is important to note that (\ref{eq:F_B}) assumes that a constant area is being submerged. While a linear force was used in this work, nonlinear buoyant forces have been considerd by \cite{Lawson2014}.
\subsection{Drag Force $F_D$}
The viscous drag is modeled using a quadratic Morison drag form \cite{Chakrabarti2005,Zurkinden2014}. The drag in the heave direction is given by
\begin{align}
F_D= -\frac{1}{2} \rho A_p C_D V|V|, \label{eq:F_D}
\end{align}
where $V$ is the relative velocity between the object and the water particles surrounding it. The density of the fluid is given by $\rho$, and $A_p$ is the projected area of the object in the heave direction. A signed velocity square term is computed by taking the product of velocity, $V$, and the absolute value of the velocity, $|V|$. Finally, the drag coefficient is given by $C_D$. A discussion on the treatment of the viscous drag force as well as an identification method can be found in \cite{Davis2018} A small linear damping force $bV$ is included to compensate for the fluid shear forces acting on the body and linear damping in the mooring lines. The coefficient $b$ is based on the experimental response of body at low velocities.
\subsection{Radiation Force $F_R$}
\indent The force acting on the WEC that results from radiated waves is calculated in the time domain through the use of the convolution integral as illustrated in \cite{Cummins1962}. A state space realization of the convolution integral is created as an efficient method of computing the radiation force. The state space representation of the radiation force can be formed as
\begin{align}
\dot{X}_r(t)=& A_rX_r(t)+B_r\dot{y}(t),\\ 
F_R =&-C_rX_r(t)-D_r\dot{y}(t), \label{eq:F_R}
\end{align}
where $X_r$ is the state variable of the realization. The state matrices $A_r,B_r,C_r,$ and $D_r$ are a non-unique linear realization of the convolution integral and are generated from the impulse response function \cite{Kristiansen2005} of the system. A more detailed description of the Fourier transform inversion and computational considerations of describing the radiatoin force as a state space system is given in \cite{Davis2018}. The work of \cite{Perez2008} states that for the radiation realization to have a causal input-output relationship, $D_r=0$.
\subsection{Excitation Force $F_{ex}$}
The wave excitation force can be calculated using the excitation force Impulse Response Function (IRF), $F_{IRF}(t)$, and the wave surface elevation, $y_w$, using the convolution integral, 
\begin{align}
F_{ex}(t)=\int_{-\infty}^{\infty}F_{IRF}(\tau)y_w(t-\tau)d\tau. \label{eq:F_ex}
\end{align} 
This convolution is used to compute the true value of the excitation force for this work. The time range of the numerical integration is truncated to be longer than the active response of the excitation IRF which is determined by applying the inverse Fourier transform on the frequency dependant excitaiton force calculated from a BEM \cite{Falnes2002}.

\section{Available Data} \label{sec:experiment}
\indent Experimental data was made available by the Oregon State University Northwest National Marine Renewable Energy Center's Marine and Hydrokinetics Open Data Project \cite{Opendata2017}. The data set that is used in this work was a physical model test of a 1:10 scale floating power system at the COAST lab at Plymouth University \cite{Bosma2015}. The deep water tests described in \cite{Bosma2015} is used to determine the parameters in the equations of motion.\\
\indent The mooring study described in \cite{Bosma2015} used a wave tank to compare several mooring configurations of two bodies in water. Of these two bodies, the scaled Floating Power System (FPS) is analyzed in this current work as it provides a very similar geometry to the float of a point absorbing WEC. For this work a subset of experimental runs were chosen. Seventeen different panchromatic runs using an elastic ``o-ring" as a taught mooring are used to test the Kalman filter estimation algorithm in section \ref{sec:results}. Table \ref{tab:runs} shows the significant wave height, $H_s$ and peak period, $T_p$ of the incident wave spectra. Each data set had roughly six-hundred seconds of data collected. \\
\begin{table}
\centering
\begin{tabular}{|c|c|c|}
\hline 
Run No. & $H_s$ mm & $T_p$ [s] \\ 
\hline 
1 & 175 & 1.42 \\ 
\hline 
2 & 50 & 1.90 \\ 
\hline 
3 & 175 & 1.90 \\ 
\hline 
4 & 300 & 1.90 \\ 
\hline 
5 & 50 & 2.37 \\ 
\hline 
6 & 175 & 2.37 \\ 
\hline 
7 & 300 & 2.37 \\ 
\hline 
8 & 50 & 2.69 \\ 
\hline 
9 & 175 & 2.69 \\ 
\hline 
10 & 300 & 2.69 \\ 
\hline 
11 & 50 & 3.16 \\ 
\hline 
12 & 175 & 3.16 \\ 
\hline 
13 & 175 & 3.95 \\ 
\hline 
14 & 175 & 4.74 \\ 
\hline 
15 & 175 & 5.53 \\ 
\hline 
16 & 375 & 2.06 \\ 
\hline 
17 & 375 & 2.37 \\ 
\hline 

\end{tabular} 
\caption{List of experimental runs from \cite{Bosma2015} that are considered in this work with corresponding significant wave height, $H_s$, and peak period, $T_p$}
\label{tab:runs}
\end{table}
\indent Incident wave elevation and mooring line forces were measured at 128Hz and motion data was recorded at a sampling frequency of 200Hz. Irregular waves were generated using the Bretshneider spectrum.
%\begin{figure}[h!]
 %   \begin{center}
%       \includegraphics[width= 85mm]{experPic.png}
%    \end{center}
 %   \caption{The 1:10 scale Floating Power System configured with compliant moorings and umbilical cable described in \cite{Bosma2015} and made available by the Northwest National Marine Renewable Energy Center's Marine and Hydrokinetics Open Data Project \cite{Opendata2017}.}
 %   \label{FPS}
%\end{figure}
\subsection{Reproducing Numerical Work}
\indent Detailed descriptions of the experiments along with wave tank schematics and dimensioned drawings can be found in the Marine and Hydrokinentics Open Data Project \cite{Opendata2017}. Key paramters used in the WEC models not found in \cite{Bosma2015} or \cite{Opendata2017} can be found in \cite{Davis2018}. Frequency dependent terms, such as added mass, radiation damping, and wave excitation are calculated using the BEM code NEMOH \cite{NEMOH}. The scripts used for this paper to execute NEMOH and all other figures will be made available on GitHub \cite{DavisGitHub}.

\section{Continuous-Discrete Extended Kalman Filter} \label{sec:algorithm}
The Kalman filter incorporates measured outputs, a model, and Gaussian error description to determine a statistically significant estimate of the state dynamics.

A nonlinear continuous time model with discrete time experimental data points can be observed using a Continuous-Discrete Extended Kalman Filter. The EKF is a nonlinear adaptation of the Kalman filter which observes a dynamic system in which zero-mean Gaussian white noise is assumed in both the dynamics and measurement. The model that is observed by the extended Kalman filter is described by
\begin{align}
\label{eq:xdot}
\dot{\textbf{x}}(t)&=\textbf{f}(\textbf{x}(t),\textbf{u}(t),t)+G(t)\textbf{w}(t) \\ \label{eq:EOM}
\tilde{\textbf{y}}(t)&=\textbf{h}(\textbf{x}(t),t)+\textbf{v}(t)
\end{align}
where $\textbf{f}$ is the nonlinear state dynamics, $\textbf{h}$ is the nonlinear output function, $\textbf{G}(t)$ is a vector that describes the noise of each state, $\textbf{w}(t)$ is zero-mean Gaussian white noise with a variance $Q(t)$, and $\textbf{v}(t)$ is zero-mean Gaussian white noise with a variance of $R(t)$. The states, $\textbf{x}(t)$, and output, $\tilde{\textbf{y}}(t)$, are written in bold font to symbolize vector quantities. The tilde symbol, $\tilde{\bullet}$, modifying the output, $\textbf{y}$, represents that this is an experimentally measured quantity. The accuracy of the Kalman filter relies on knowledge of the process and measurement noise terms. Correct quantification of the process noise is essential for the EKF to function effectively. For discussions concerning the determination of the process noise covariance see \cite{Crassidis2011,Mehra1970}. Since the EKF deals with nonlinear systems that must be integrated to determine state dynamics, reasonable initial conditions must be used to ensure filter stability \cite{Crassidis2011}. 

The Kalman filter will iterate through the gain, update, and propagation steps for each discrete time in order to estimate the states of the system. The gain step calculates a matrix of gains that is used to correct a previous step's estimates based on the covariance of each state. The update step uses the calculated gain to adjust the covariance of each state and estimated value based on the difference between estimated and measured states. Finally the propagation step calculates the covariance and state estimates of the next time step.\\
\indent Details on the theory of the continuous-discrete EKF can be found in \cite{Crassidis2011} and the implementation of the algorithm on a WEC application can be found in \cite{Davis2018}. For this implementation of the EKF the Joseph stabilized form of the covariance update equation is used, and to ensure the symmetry of the covariance matrix is enforced by taking the element-wise average of the covariance and its transpose at each step. In order to use the Kalman filter for estimation the model must be modified to provide the estimated parameter as a function of the observed states as described below.
\subsection{Direct Identification Estimation}
\indent To directly identify a force or parameter using the EKF the variable in question is appended as an extra state with no dynamics. Each state in the dynamics of a Kalman undergoes a ``random walk'' to determine the best estimate of the state variable at each time step \cite{Crassidis2011}. The nonlinear dynamics in (\ref{eq:xdot}), described in section \ref{sec:wecModel}, when reduced to a first order system is,
\begin{align}
\begin{bmatrix} \dot{x}_1 \\ \dot{x}_2 \\ x_r \end{bmatrix} = \begin{bmatrix} x_2 \\ \frac{1}{M_v}F(x_1,x_2,F_{ex},\dot{y},C_rx_r) \\ A_rx_r+B_r\dot{y} \end{bmatrix}
\end{align}
where $x_r$ is a vector of radiation states, $x_1$ and $x_2$ are the first order variables for position and velocity respectively, and both $F_{ex}$ and $\dot{y}$ are inputs to the system representing wave excitation force and water velocity, respectively. The estimated wave excitation force can be used to provide a functional estimate of the water velocity surrounding the WEC as described in section \ref{sec:estWater}.\\
\indent The inclusion of $F_{ex}$ as an estimated force necessitates the inclusion of another state in the dynamics as,
\begin{align}
\begin{bmatrix} \dot{x}_1 \\ \dot{x}_2 \\ \dot{x}_3 \\ \dot{x}_r \end{bmatrix} = \begin{bmatrix} x_2 \\ \frac{1}{M_v}F(x_1,x_2,x_3,\dot{y},C_rx_r) \\ 0 \\ A_rx_r+B_r\dot{y} \end{bmatrix} \label{eq:fexAppend}
\end{align}
where $F_{ex}$ is now the third state of the system estimated by the EKF. The inclusion of a state with zero dynamics means that the ``random walk'' that takes place during the gain and propegation steps of the Kalman filter is responsible for the changes in the estimated value. Equation (\ref{eq:fexAppend}) assumes that the time derivative of the excitation force has the form $\dot{F}_{ex} = gw(t)$, where $g$ is a single term of $G(t)$ in (\ref{eq:xdot}) and $w(t)$ is a Gaussian zero-mean white noise signal. The EKF is then modeling the excitation force simply as noise where the current value of $F_{ex}$ does not depend on previous estimates. The choice of the process noise $w(t)$ for the estimated force is crucial to success of the estimation algorithm, the selection of noise is discussed in \cite{Crassidis2011} and the choice used in this work is described in section \ref{sec:noise}.\\
\indent The direct identification method only adds a single state to the dynamics of the system and as such is the most computationally simple identification method used in this work. An additional initial condition is necessary with the addition of any state is required, however precise determination of this initial condition is not necessary as it is expected that there be an initial error in the dynamics. This method of identifying the excitation force does not require information on the wave climate, however a sensitivity analysis of the process noise corresponding to the identified state should be performed. An example where a large process noise for an identified state is able to identify a time-varying viscous drag coefficient is shown in \cite{Davis2018}.
\subsection{Disturbance Model Estimation} 
An alternative method of estimating the wave excitation force is to model the incident wave force as a disturbance to the WEC dynamics \cite{LingThesis2015}. Ocean waves and excitation forces are typically modeled as harmonic oscillator, or the sum of a large number of sinusoids, with parameters determined by a wave energy spectrum and randomized phase \cite{Folley2016}. The disturbance, as described by \cite{Ling2015}, will be modeled using a similar approach. The regular wave representation of the excitation force is given as
\begin{align}
F_{ex}=\mathbb{R}\left[ \frac{H}{2}F_X(\omega_r) e^{i(\omega_rt) } \right], \label{eq:FexRegular}
\end{align}
where $\mathbb{R}$ represents the function to extract the real part of the argument, $H$ is the peak to peak wave height, and $F_X(\omega_r)$ is the magnitude of the frequency domain excitation force for a given frequency, $\omega_r$. The imaginary unit is donated by $i$, and $t$ is the time in seconds. This excitation force can be generalized to irregular waves by 
\begin{align}
F_{ex}=\mathbb{R}\left[\int^\infty_0\sqrt{2 \Delta\omega S(\omega)} F_X(\omega) e^{i(\omega t+\phi (\omega))}d\omega \right],\label{eq:FexIrreg}
\end{align}
where $S(\omega)$ is the power of the wave spectrum as a function of frequency. The variable $\Delta\omega$ is the discrete distance separating each frequency and $\phi$ is vector of randomly generated phase angles in radians \cite{WECsim}. \\
\indent Equation (\ref{eq:FexIrreg}) shows an infinite integral over a range of frequencies which is typically modeled a discrete sum \cite{Folley2016}. By looking at a single frequency component of the wave excitation force a description of the excitation force dynamics can be shown as,
\begin{align}
f_{ex}&=sin(\omega t) \nonumber \\ 
\ddot{f}_{ex}&=-\omega^2sin(\omega t).\label{eq:ddotfex}
\end{align} 
The differential equation describing the excitation force can be implemented in the first order system of equations by 
\begin{align}
\begin{bmatrix} \dot{x}_1 \\ \dot{x}_2 \\ \dot{f}_{ex} \\ \ddot{f}_{ex} \\ \dot{x}_r \\ \end{bmatrix} = \begin{bmatrix} x_2 \\ \frac{1}{M_v}F(x_1,x_2,f_{ex},\dot{y},C_rx_r) \\ \dot{f}_{ex} \\ -\omega ^2\dot{f}_{ex} \\ A_rx_r+B_r\dot{y} \end{bmatrix}, \label{eq:distsize1}
\end{align}
where $f_{ex}$ and $\dot{f}_{ex}$ are now states of the dynamic system. The above expression describes a system where a single excitation force signal is estimated, similar to the direct identification method. However, (\ref{eq:distsize1}) enforces that the estimated excitation force be a solution to the differential equation given in (\ref{eq:ddotfex}).\\
\indent This excitation force estimation method can be generalized to use multiple wave frequencies by using
\begin{align}
\begin{bmatrix} \dot{x}_1 \\ \dot{x}_2 \\ \dot{f}_{ex} \\ \ddot{f}_{ex} \\ \dot{x}_r \\ \end{bmatrix} = \begin{bmatrix} x_2 \\ \frac{1}{M_v}F(x_1,x_2,\Sigma \bar{f}_{ex},\dot{y},C_rx_r) \\ I\dot{\bar{f}}_{ex} \\ -\Omega ^2\dot{\bar{f}}_{ex} \\ A_rx_r+B_r\dot{y} \end{bmatrix} \label{eq:distsizeN}
\end{align}
for the dynamics. The bar over the excitation force denotes a vector of disturbance forces. The excitation force becomes $F_{ex}=\Sigma \bar{f}_{ex}$ to denote the sum of $N$ excitation force terms that each oscillate at a different frequency. The term $\Omega$ denotes the diagonal matrix containing each of the $N$ frequencies corresponding to the entries of $\bar{f}_{ex}$. The sum of each excitation force frequency allows for the excitation force to be modeled in a way that realistically describes the time derivative behavior of the wave excitation force. This method however requires knowledge of the wave climate for an appropriate choice of the disturbance wave frequencies. The choice of these frequencies is discussed in section \ref{sec:EEA}.\\
\indent Just as with the direct identification method, each of the new states that are included in the dynamics are subject to a ``random walk'' that is influenced by the process noise covariance in the gain and propagation steps of the Kalman filter. The disturbance estimation method becomes nonlinear if any of the frequencies in $\Omega$ are varied, where any frequencies would then be appended to the dynamics as an extra state \cite{Ling2015}. For this work the frequencies remain constant. 

The initial conditions of the excitation terms should satisfy the differential equations given by (\ref{eq:ddotfex}). These initial conditions act as the terms that dictate the phase of the excitation force terms, however after a small amount of simulation time the Kalman filter is able to adjust to reasonable errors given by the initial conditions. As a result, once the Kalman filter is properly set up then nearly any physical initial condition can be used to start the estimation. 

The disturbance model introduces two states for every disturbance frequency that is used in the dynamics which is a significant increase in computational complexity over the direct identification method. However, the use of the equal energy approach when selecting wave frequencies provides an effective way to model the incident wave with only few frequencies. This enables the use of the disturbance model to more realistically describes the wave excitation force.
\subsubsection{Equal Energy Approach}\label{sec:EEA}
\begin{figure}[h!]
    \begin{center}
       \includegraphics[width= 85mm]{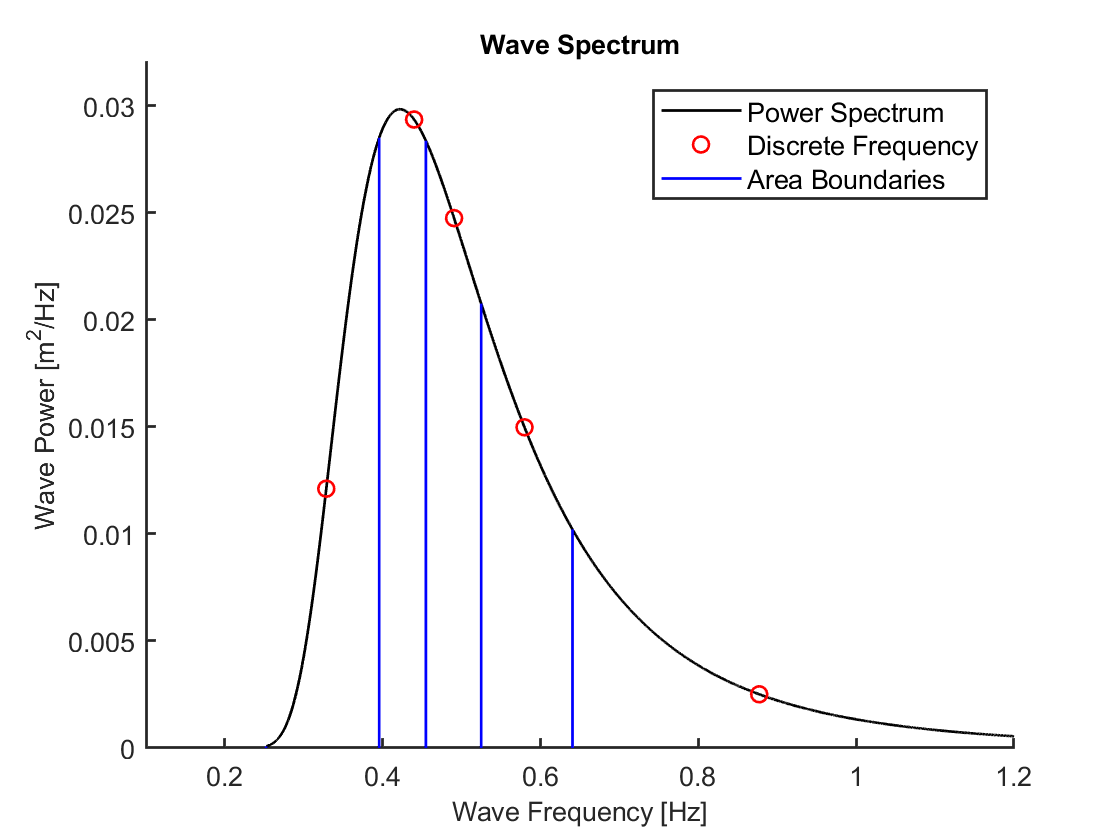}
    \end{center}
    \caption{The wave spectrum of run 17 generated by the Bretschneider Spectrum. Red circles indicate the frequencies selected for use in the disturbance method and blue lines are the divisions between the equal energy areas of the spectra.}
    \label{EEA}
\end{figure}

The implementation of the disturbance method requires the choice of $N$ frequencies to comprise the wave excitation force. The typical approach to representing irregular waves uses many frequencies, on the order of 10,000 sinusoids \cite{Nguyen2017}, however this number would be computationally impractical for implementation in a Kalman filter. The wave frequencies used to construct an irregular wave can be found in several ways as described in \cite{OrcaFlex}. Some examples include an arithmetic progression, which selects wave components that are equally spaced in frequency, and a geometric progression, which selects wave components that have a constant ratio of wave frequency. Both of these methods typically require large numbers of waves to realistically represent an irregular wave. In contrast, the equal energy method can be used to realistically represent an irregular wave with relatively few frequency components, and is mentioned in \cite{Folley2016}. The basic premise of this method is to select wave components such that each frequency represents an equal amount of spectral energy. Since this energy is calculated by considering the area under the energy spectrum, this method is sometimes referred to as the equal area approach.\\
\indent Figure \ref{EEA} shows the wave energy spectra of the 17th run of the available experimental data described in section \ref{sec:experiment}. The significant wave height of the generated wave was 375 mm and the peak period was 2.37 seconds. The vertical blue lines show the divisions between each equal energy area of the spectra and the red circles indicate the power and frequency of the components that were selected to form the disturbance equations.\\
\indent The divisions are created by determining the area under the energy spectra and dividing this by the number of desired frequency components, which was in this case 5 components, to determine the area of each division. A threshold for minimum power is required otherwise the very high and very low frequency components would artificially bias the selection towards frequencies with very little energy. Once the equal energy areas are determined a single frequency with the average power is selected from each equal energy area to provide the average power for that division. This method of determining the disturbance frequencies is effective for selecting wave frequencies and initial estimations of each components magnitude while only requiring the peak period and significant wave height of the incident wave spectra. \\
\subsection{Implementing the EKF}
\indent In order to implement the extended Kalman filter a few key considerations were required prior to successful estimation of the excitation force. First, and often the most difficult part of any Kalman filter, is determining the process noise covariance given in section \ref{sec:noise}. The other barrier to the implementation of the nonlinear equation of motion is that in typical WEC models both the excitation force and incident water velocity are unknown, therefore an approximation of the wave velocity is required. The method used in this work for estimating the water velocity is described in section \ref{sec:estWater}. 
\subsubsection{Process Noise} \label{sec:noise}
\indent The process noise covariance is the product of $G(t) \textbf{w}(t)$ that describes the error in the dynamics of the system. The diagonal entries of the symmetrical matrix $G(t) \textbf{w}(t)$ are variances of the noise in each state and non diagonal entries quantify the correlation between the states. A typical assumption for the development of a Kalman filter is to assume that no cross-correlation exists between states \cite{Crassidis2011}. For the equation of motion the radiation states are excepted from having a process noise variance as the error in the radiation force is included in the error for the position and velocity states \cite{Davis2018}.\\
\indent The process noise for each state was found by simulating the response of the model to the inputs determined from experimental measurements. The convolution formulation of the excitation force in (\ref{eq:F_ex}), is used to determine the wave excitation force that the float experienced during each experimental run, and the water particle velocity is known from measurements of the incident wave \ref{sec:experiment}. The dynamics of the float are simulated using the experimental measurements and the states are compared to the measured states from \cite{Bosma2015}. The process noise describes the error between the simulated states and the experimental states, so the variance of this simulation error is calculated for each run and used as the process noise covariance in the EKF.\\
\indent Recall that we are considering various experiemental runs from \cite{Bosma2015}, and attempting to determine the wave excitation force for each run. Since the process noise is a variable, with dimensions, that will vary based on the incident wave it becomes necessary for each run to have its own process noise. A typical Kalman filter implementation requires a significant amount of tuning to provide efficient estimation is achieved \cite{Crassidis2011} and in some cases optimization algorithms are used to identify a ``best'' choice of the process noise terms \cite{LingThesis2015}. For this work, a reasonable value of process noise for the excitation force was chosen for each separate sensitivity analysis as changes in sampling rate, model complexity, and measurement noise all change the robustness of the EKF to changes in the estimated state process noise.
\subsubsection{Estimating Water Velocity} \label{sec:estWater}
The estimation of the water velocity of the incident wave enables the calculation of the Morison drag force as the relative velocity is needed between the body and the surrounding water particles. With this estimation of the velocity it becomes possible to observe the model described in section \ref{sec:wecModel} using an EKF. It is important to note that this nonlinear time domain model is well represented as a generic model for point absorbing WECs \cite{Bhinder2015}.\\
\indent The estimation of the water particle velocity relies on a regular wave representation of the wave excitation force using (\ref{eq:FexRegular}). When simulating the excitation forces in irregular waves using (\ref{eq:F_ex}) it is necessary to incorporate the time delay and the frequency dependent magnitude of the excitation force, however in order to approximate the incident wave position the magnitude is assumed to have a small phase lag and a magnitude equal to the magnitude of $F_X(\omega_p)$, where $\omega_p$ is the peak frequency of the exciting wave.\\
\indent By using the linear relationship of (\ref{eq:FexRegular}) between the wave elevation and excitation force, the wave elevation can be approximated by 
\begin{align}
y(t) = \frac{F_{ex}(t)} {||F_X(\omega_p)||_2}
\end{align}
where $F_{ex}(t)$ is the estimated excitation force and $||F_X(\omega_p)||_2$ is the L2-norm, Euclidean norm, of the excitation force at the peak frequency of the incident wave. This norm is required as each entry of $F_X(\omega)$ is an imaginary number that is used to represent the magnitude and phase of the wave excitation force. The wave velocity can then be computed using a finite difference method based on the previous estimations of the water elevation. This simple method allows for the full implementation of the Morison drag equation with the time domain model discussed in section \ref{sec:wecModel}. The estimation results of the water particle velocity are given in section \ref{sec:results}.

\section{Estimation Results} \label{sec:results}
The two estimation methods presented in this work were tested using the experimental data described in section \ref{sec:experiment}. The discrete measurements used in the EKF are the heave position and velocity, and the estimated excitation force is compared to the excitation force calculated using the non-causal convolution integral formulation in (\ref{eq:F_ex}). This convolution integral relies on knowledge of the time history of the incident wave as well as the impulse response of the buoy in the heave direction, both of which are known before estimation as a result of the work in \cite{Bosma2015}. The experimental values of water velocity is determined by taking the time derivative of the measured water surface elevation. 

Each estimation run is evaluated using the Normalized Mean Square Error (NMSE), given by the expression
\begin{align}
NMSE=1-\frac{||s(t)-\tilde{s}(t)||_2}{||s(t)-\bar{s}(t)||_2} \label{eq:NMSE}
\end{align} 
%NMSE=1-\frac{\Sigma(\tilde{s}(t)-s(t))^{2}}{\Sigma( s^2(t) - \bar{s}(t))^2}, \label{eq:NMSE}

where $s(t)$ is the time domain reference signal, the bar notation denotes the mean value, $\tilde{s}(t)$ is the estimated signal, and the notation $||\bullet||_2$ represents the L2-norm of a vector. The NMSE goodness of fit parameter is a scalar with a value less than one. Alternatively the goodness of fit can be represented as a percentage as shown in \cite{Nguyen2017}. In this work the value will be represented as a scalar where $NMSE=1$ would denote perfect estimation of the reference signal \cite{MATLAB_GOF} . In the case of this work both the incident wave velocity and excitation force are evaluated using (\ref{eq:NMSE}). 

\begin{figure}[h!]
    \begin{center}
       \includegraphics[width= 85mm]{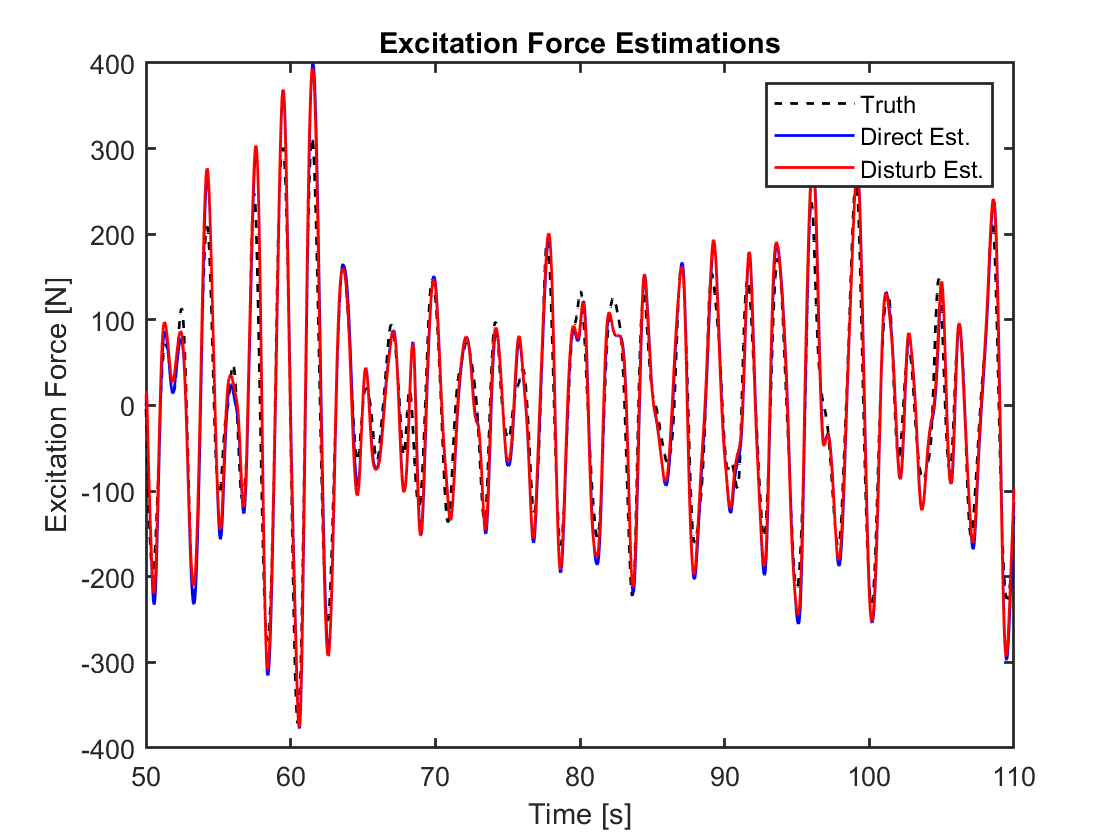}
    \end{center}
    \caption{Time domain plot of estimated excitation force for a sixty second window using run number 7.} 
    \label{TDest}
\end{figure}

\begin{figure}[h!]
    \begin{center}
       \includegraphics[width= 85mm]{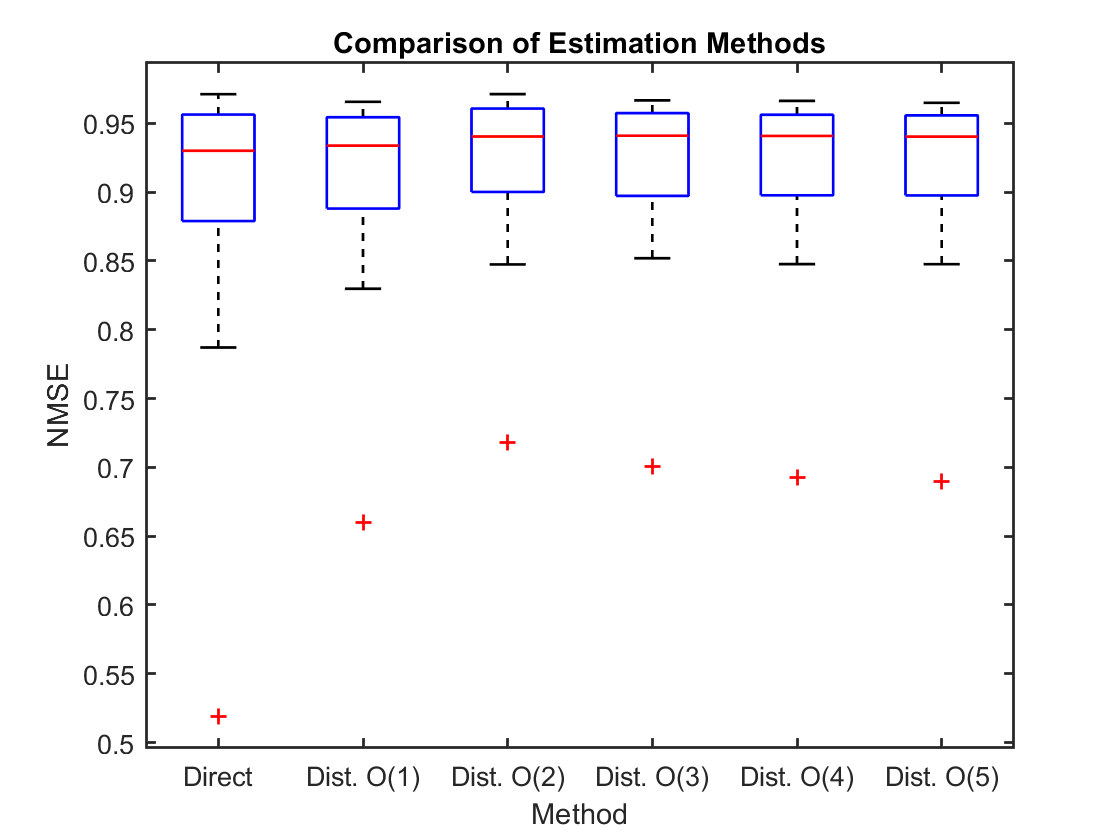}
    \end{center}
    \caption{The comparison the direct estimation method and five variations of the disturbance method for wave excitation force. O(*) refers to the number of sinusoids used in the disturbance method. Outliers are denoted with a red $+$.} 
    \label{NMSEmethods}
\end{figure}

Figure \ref{TDest} shows the time domain estimation of the wave excitation force using both the direct identification method and the third order disturbance estimation method, where \textit{third order} signifies that the disturbance models is a sum of three excitation force frequencies, for the incident wave. The reference signal that was calculated from experimental measurements is shown in black, and the results from the direct identification and the disturbance estimation methods are shown in blue and red, respectively. The third order disturbance estimation is shown as it provides a representative example of the performance of the disturbance estimation method while also being directly comparable to the work of \cite{Ling2015}. This figure shows that some of the very sharp changes in the excitation force are not captured by the estimation algorithm, however it is apparent that the estimation algorithm results in a excitation force signal that represents well the true signal. For this estimation run the significant wave height of the incident wave was $H_s=300$ mm and the peak period was $T_p=2.37$ s. The NMSE of direct and disturbance estimations are $0.914$ and $0.927$, respectively. In order to test the estimation algorithms in a variety of wave climates the excitation force estimation was performed for the seventeen different wave climates described in section \ref{sec:experiment}, the results of which are shown in Fig. \ref{NMSEmethods} which is described below.

\begin{figure}[h!]
    \begin{center}
       \includegraphics[width= 85mm]{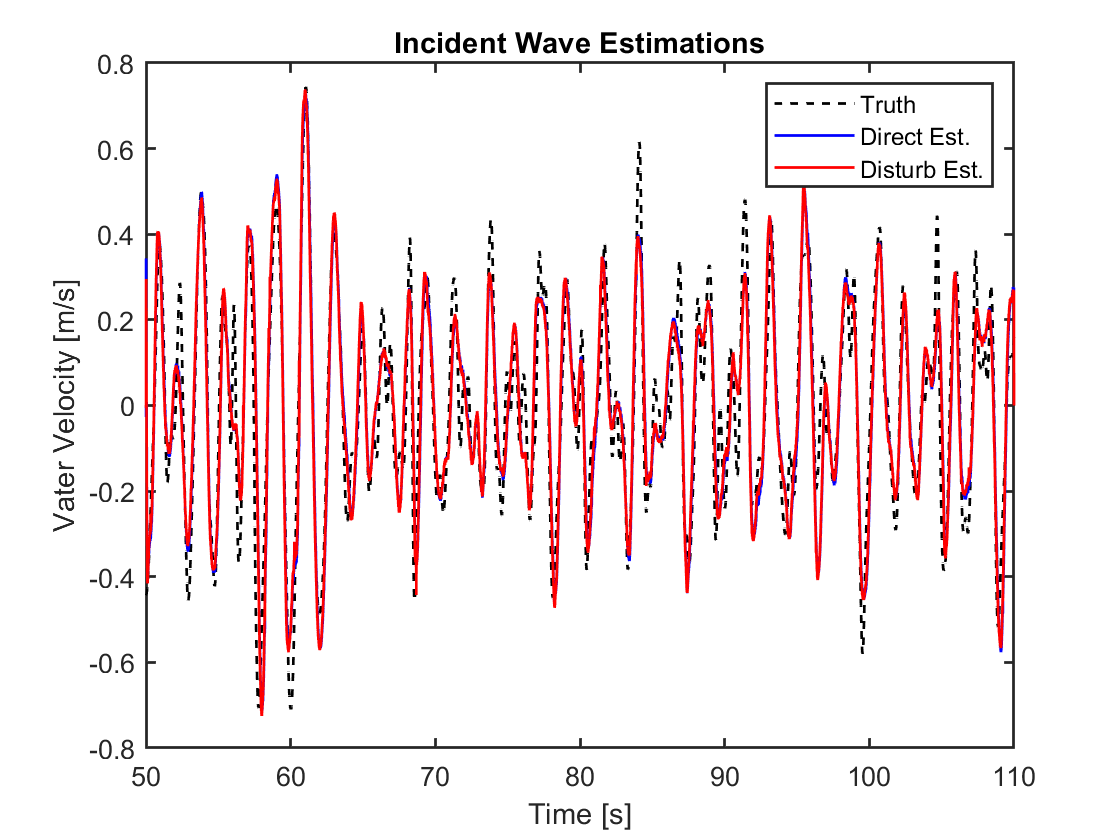}
    \end{center}
    \caption{Time domain plot of estimated water velocity for a sixty second window using run number 7.} 
    \label{TDvel}
\end{figure}

The estimation of the incident wave water velocity is shown in Fig \ref{TDvel} for the run 8 in table \ref{tab:runs}. The water velocity results from the direct estimation implementation and the disturbance estimation implementation are once again shown in blue and red, respectively. The method by which the incident water velocity is estimated is described in section \ref{sec:estWater}. It is not difficult to tell by visual inspection that estimation of the water velocity is not as accurate as the estimation of the excitation force. This is reflected in the goodness of fit values, which for the direct and disturbance estimation methods results, shown in Fig. \ref{TDvel}, are $0.825$ and $0.831$, respectively. The approximation used to describe the velocity as a function of excitation force is very simplistic and a more detailed formulation of water velocity as a function of excitation force using the relationship given by (\ref{eq:FexIrreg}) could be implemented. Additionally a higher order finite difference scheme could be implemented. However, as the influence of the water velocity on dissipation forces does not drive the dynamics of the WEC, this simple approximation method is sufficient for this work.

\begin{figure}[h!]
    \begin{center}
       \includegraphics[width= 85mm]{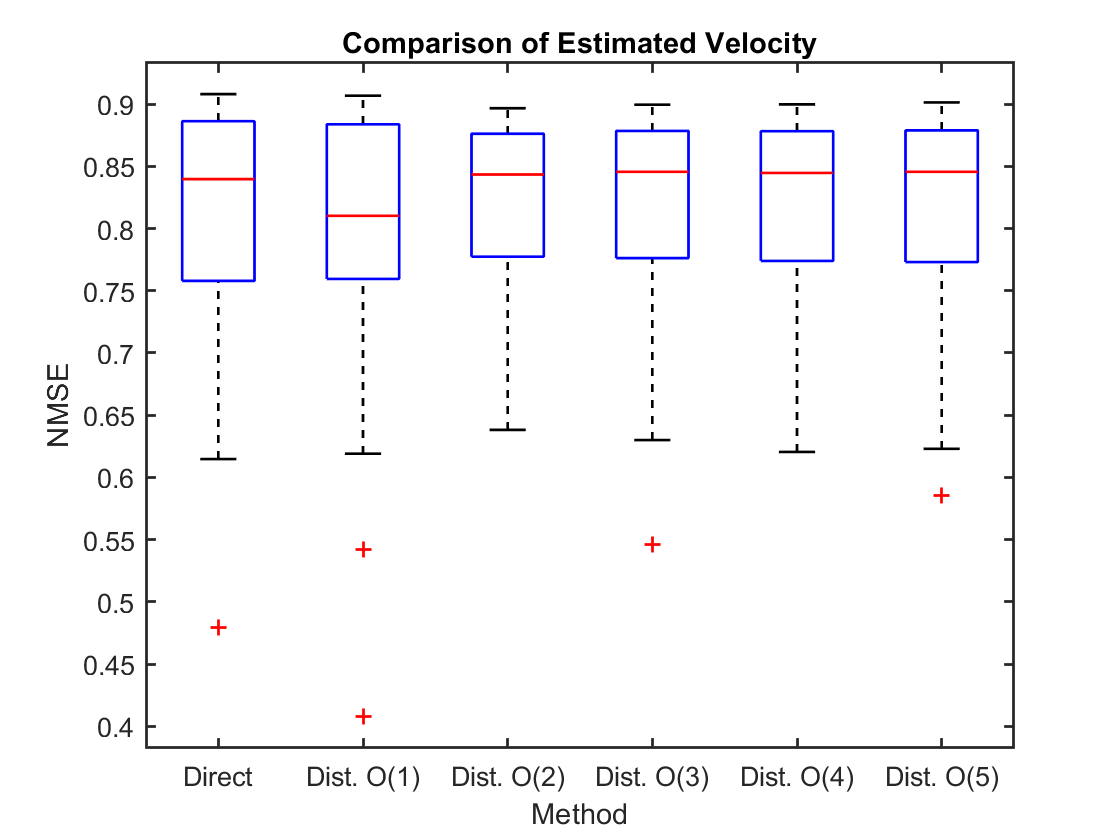}
    \end{center}
    \caption{The comparison the direct estimation method and five variations of the disturbance method for wave velocity. O(*) refers to the number of sinusoids used in the disturbance method. Outliers are denoted with a red $+$.}
    \label{NMSEmethodsVel}
\end{figure}

The NMSE of several estimation algorithms applied to each run in table \ref{tab:runs} are shown in Fig. \ref{NMSEmethods}. Recall that both of the estimation methods being presented are run seventeen times, once for each of the different wave climates being considered. The results are shown in Fig. \ref{NMSEmethods} as box and whisker plots to give an understanding of the overall performance of each method. For this comparison the direct estimation method is shown compared to the disturbance estimation method of varying orders (one excitation term to five excitation terms). It can be seen from the figure that the performance of the disturbance estimation algorithm provides the best estimation result using three excitation force components. As a result the third order disturbance estimation method will be used when analyzing the sensitivity of the estimation algorithms. The estimation of the incident wave velocity is shown in Fig. \ref{NMSEmethodsVel}. As shown in this figure the mean of the seventeen NMSEs tend to be roughly 10\% lower than the excitation force estimation shown in Fig. \ref{NMSEmethods}. 

Both figures \ref{NMSEmethods} and \ref{NMSEmethodsVel} show a single  outlier marked with a red $+$. This outlier corresponds to the first run in table \ref{tab:runs}. This high frequency wave is shown to be consistently difficult for the EKF to estimate. While it is possible to tune the parameters of the Kalman filter to better estimate the high frequency wave that is currently being shown as an outlier, the purpose of this work is to show general trends in the estimation capabilities of the EKF methods. Therefore, rather than producing the highest possible goodness of fit for a single wave climate we have focused on tuning the parameters of the Kalman filter to produce acceptable estimation for a wide range of wave climates. It is of interest to note that this first run consisted of a wave that generated relatively small excitation forces when compared to the waves from the other runs being considered. For example, the excitation force of the seventeenth run has a magnitude about four times larger than the first run. This means that the wave from the first run will have a much smaller effect on the dynamics of the buoy than any of the other runs considered in this work. Therefore, while this first run consistently has the lowest NMSE of any of the runs being considered, the waves that are consistently estimated well by the EKF have a much larger effect on the buoy dynamics. 

\begin{figure}[h!]
    \begin{center}
       \includegraphics[width= 85mm]{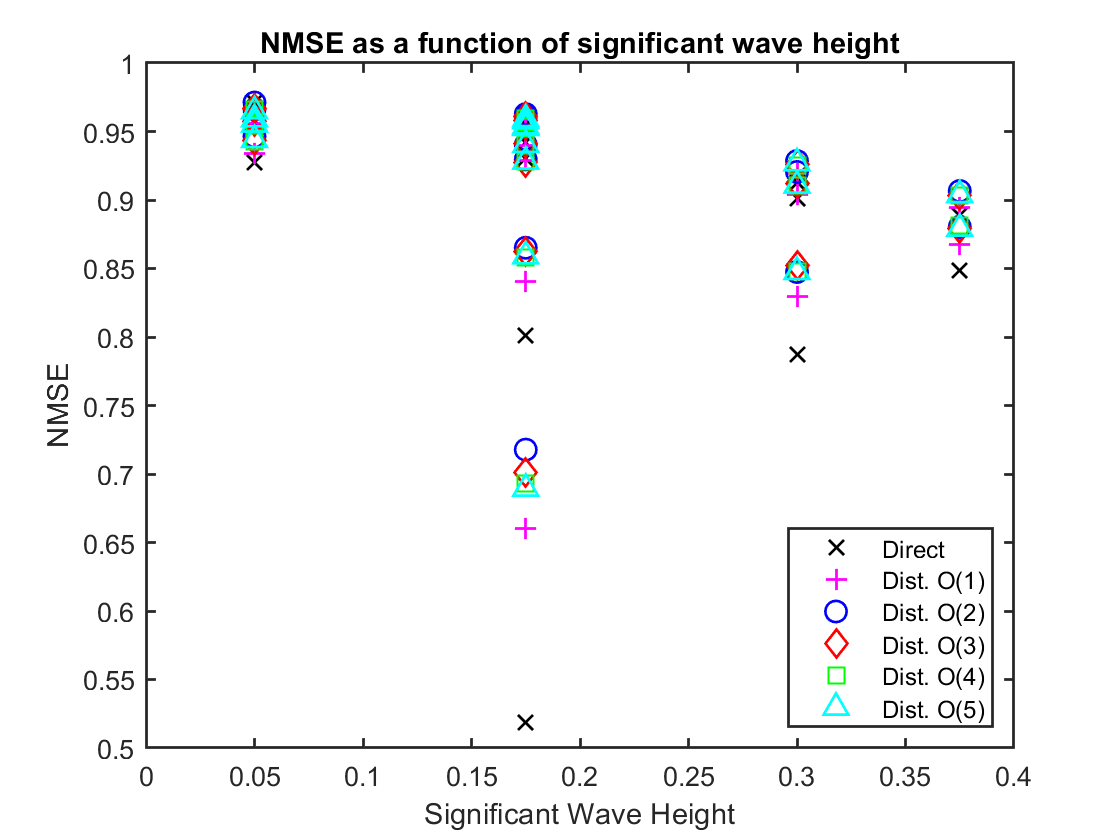}
    \end{center}
    \caption{The excitation force estimation goodness of fit of each run as a function of the peak period of the significant wave height, $H_s$. O(*) refers to the number of sinusoids used in the disturbance method.}
    \label{NMSE_Hs}
\end{figure}

\begin{figure}[h!]
    \begin{center}
       \includegraphics[width= 85mm]{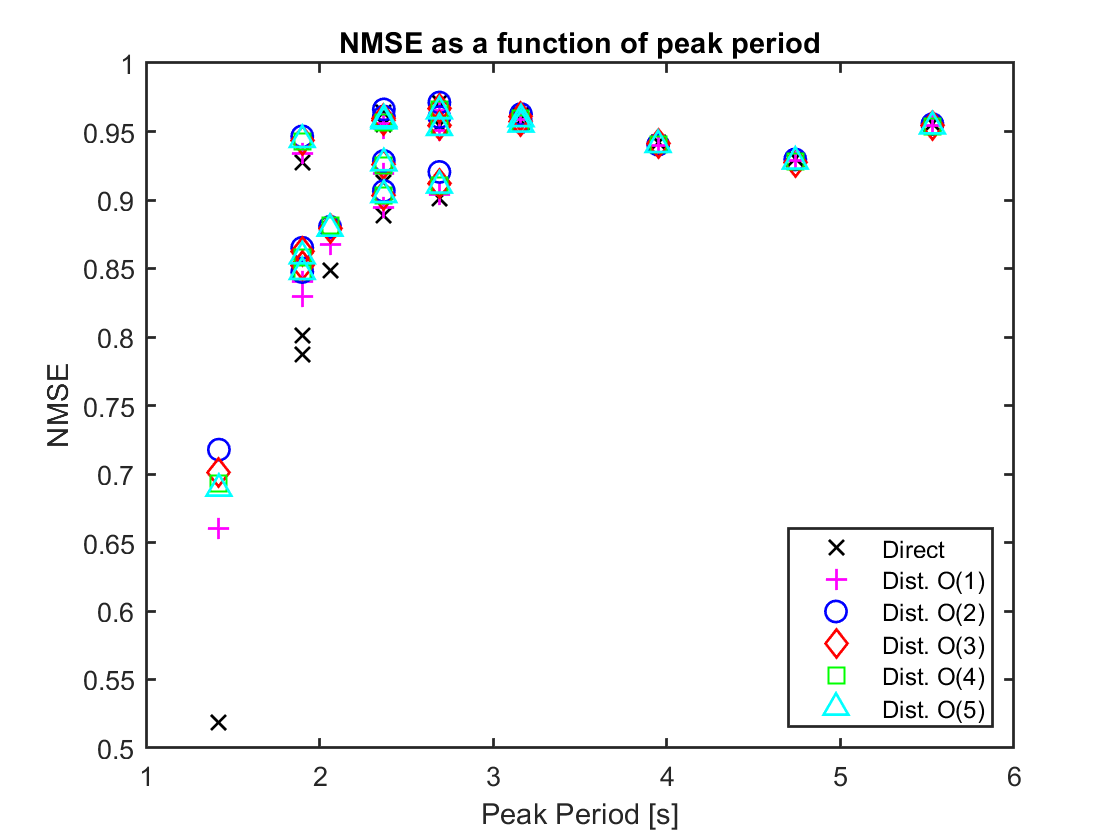}
    \end{center}
    \caption{The excitation force estimation goodness of fit of each run as a function of the peak period of the wave climate, $T_p$. O(*) refers to the number of sinusoids used in the disturbance method.}
    \label{NMSE_Tp}
\end{figure}

\begin{figure}[h!]
    \begin{center}
       \includegraphics[width= 85mm]{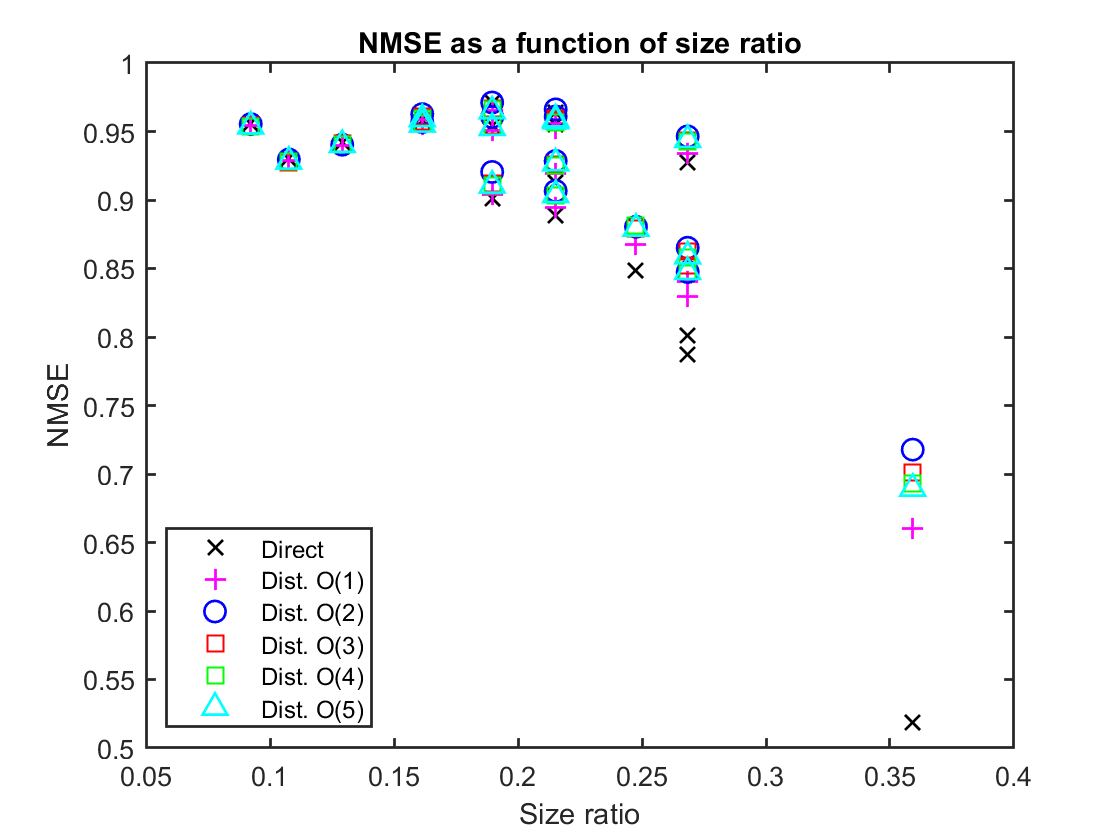}
    \end{center}
    \caption{The excitation force estimation goodness of fit of each run as a function of the dimensionless ratio between the characteristic dimension of the float and wavelength. O(*) refers to the number of sinusoids used in the disturbance method.}
    \label{NMSE_size}
\end{figure}

As each run represents a different incident wave the NMSE can be plotted against several parameters as shown in Figures \ref{NMSE_Tp}, \ref{NMSE_Hs}, and \ref{NMSE_size} which show the NMSE as a function of peak period, significant wave height, and size ratio respectively.
 
In Fig. \ref{NMSE_Hs} the NMSE is shown to not be directly influenced by the significant wave height, unlike the wave frequency and size ratio. The performance of the estimation algorithms tend to be worse in the cases where the wave is oscillating at a high frequency. Figure \ref{NMSE_Tp} shows that the lowest performing estimation runs occur at low frequencies. The size ratio is commonly used to determine  the dominant forces for modeling purposes \cite{Newman1977}. As the relative size of the float when compared to the wavelength becomes large the assumptions used in the development of the EOM for this work may be called into question as the size ratio is shown to negatively impact the NMSE of each run in Fig. \ref{NMSE_size}. 

\subsection{Sensitivity to Sampling Rate}
\begin{figure}[h!]
    \begin{center}
       \includegraphics[width= 85mm]{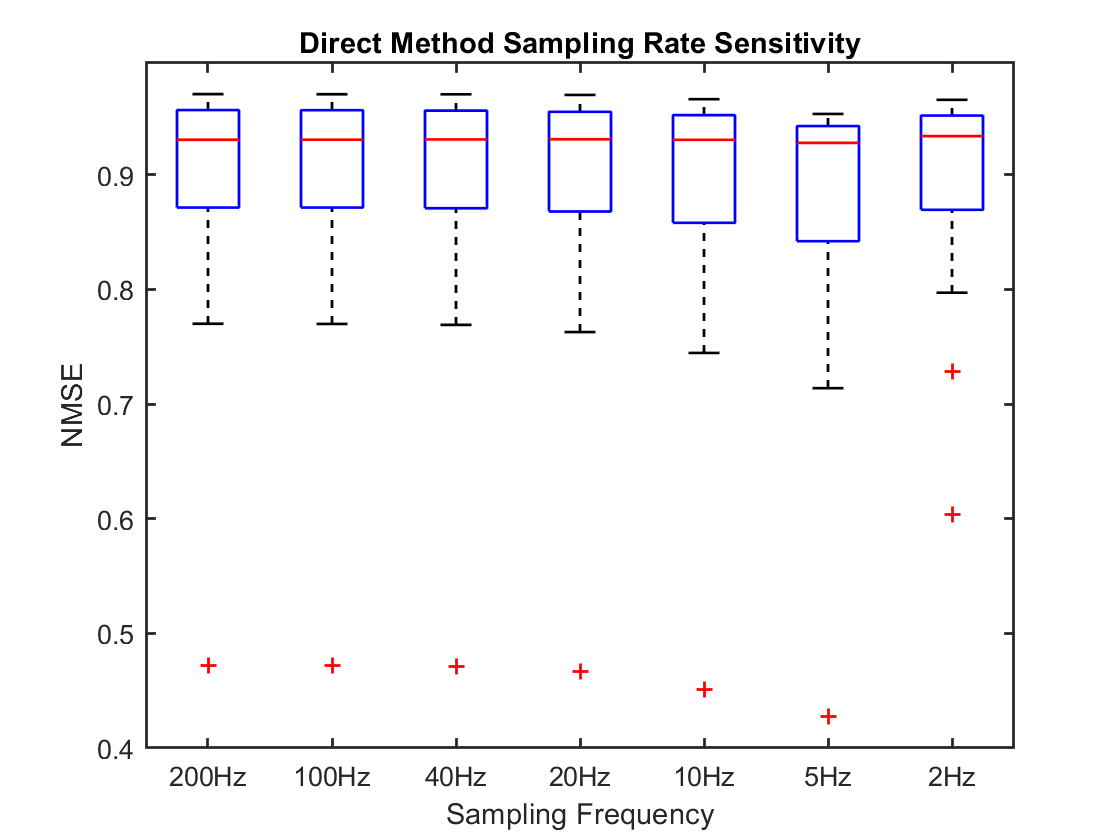}
    \end{center}
    \caption{The excitation force estimation NMSE using the direct estimation method of each run shown for various sampling frequencies. Outliers are denoted with a red $+$.}
    \label{NMSE_sampleDirect}
\end{figure}

\begin{figure}[h!]
    \begin{center}
       \includegraphics[width= 85mm]{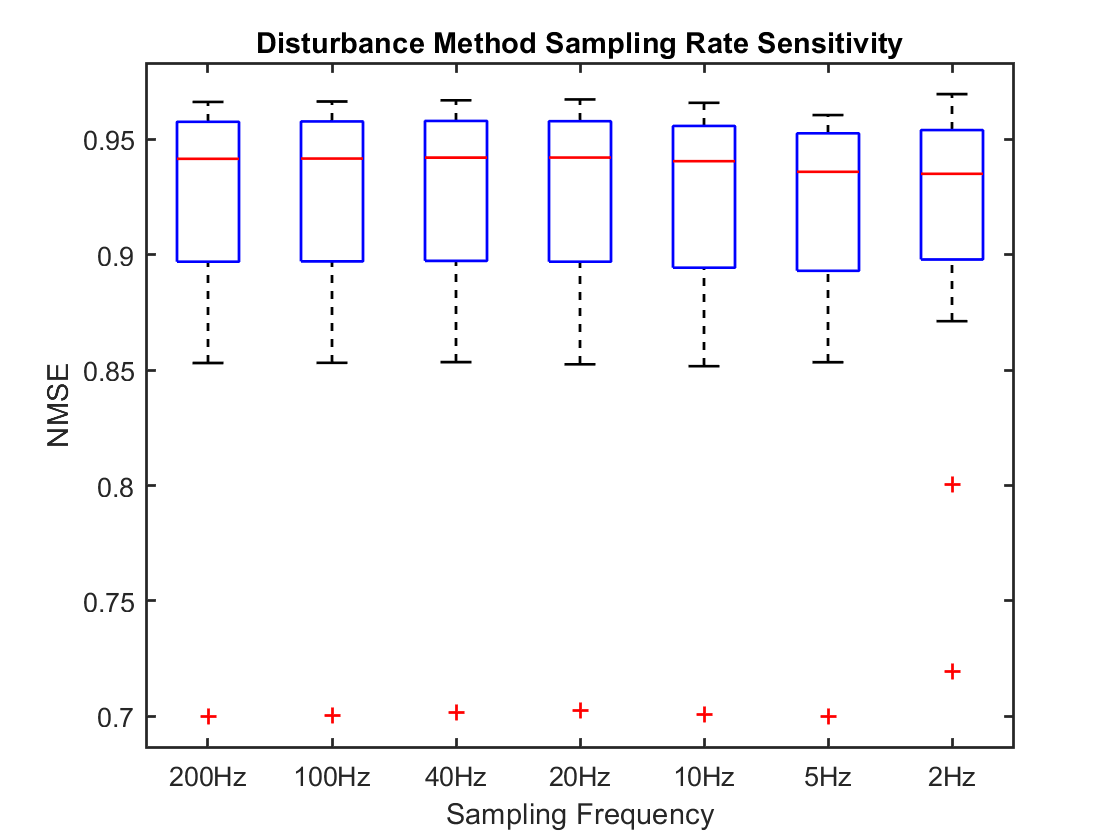}
    \end{center}
    \caption{The excitation force estimation NMSE using the third order disturbance estimation method of each run shown for various sampling frequencies. Outliers are denoted with a red $+$.}
    \label{NMSE_sampleDisturb}
\end{figure}

The motion data is originally recorded at a sampling frequency of 200Hz. A subset of the motion data is used to test the sensitivity of the estimation algorithms to sampling rate. Figures \ref{NMSE_sampleDirect} and \ref{NMSE_sampleDisturb} show the response of the direct estimation and third order disturbance estimation when the sampling rate is lowered. To select a subset of the measurements an evenly spaced number of the measurements were selected and treated as the measured data points. The resulting sampling frequencies are used as the control variable to determine the influence on the NMSE of the estimations. 

Figures \ref{NMSE_sampleDirect} and \ref{NMSE_sampleDisturb} show that even at much lower sampling rates the effectiveness of the EKF to identify the excitation force is still resulting in a NMSE greater than 0.9. The outlier corresponding to the first, high frequency, run shows a more significant performance degradation than the mean performance of the rest of the runs. It intuitively follows that a low sampling rate would adversely affect the estimation of a high frequency wave. It is important to note that while the average for the runs appears to increase at a sample rate of 2Hz the fact that an extra outlier is shown means that the overall performance for all of the runs is lower. This extra outlier in both figures  \ref{NMSE_sampleDirect} and \ref{NMSE_sampleDisturb} corresponds to run 4 which has a low period for its amplitude. The robustness of the estimation algorithms to a change in sampling rate shows promise for the potential of real time estimation for WEC deployments.

\subsection{Sensitivity to Radiation Force Order}
The radiation force is modeled using a state space realization given by (\ref{eq:F_R}) that can vary in size. Reference \cite{LingThesis2015} shows a study in which the order of the radiation realization is varied and the effect on the estimator performance is evaluated using simulated wave measurements. When generating the state space realization of the radiation force it is typical to use a coefficient of determination as shown in \cite{Davis2018}. The fourth order state space realization, used in this work, was the smallest realization that satisfied the minimum coefficient of determination threshold established in \cite{Davis2018}. However, each state used in the radiation realization increases the computational burden of the model. Therefore, a simple investigation into the effects that the order of the radiation force has on the excitation force estimation is conducted.

\begin{figure}[h!]
    \begin{center}
       \includegraphics[width= 85mm]{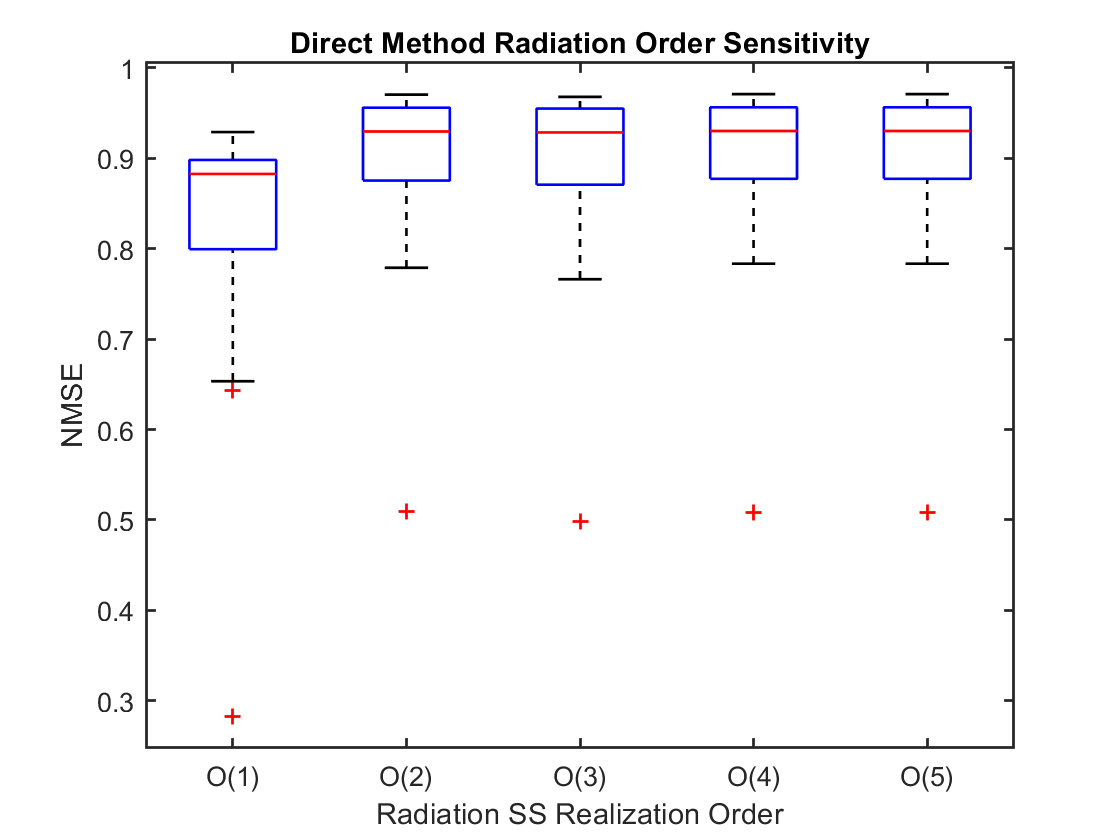}
    \end{center}
    \caption{The excitation force estimation NMSE using the direct estimation method of each run shown for various orders of the state space radiation realization. O(*) refers to the order of the state space realization of the radiation force. Outliers are denoted with a red $+$.}
    \label{NMSE_DirRad}
\end{figure}

\begin{figure}[h!]
    \begin{center}
       \includegraphics[width= 85mm]{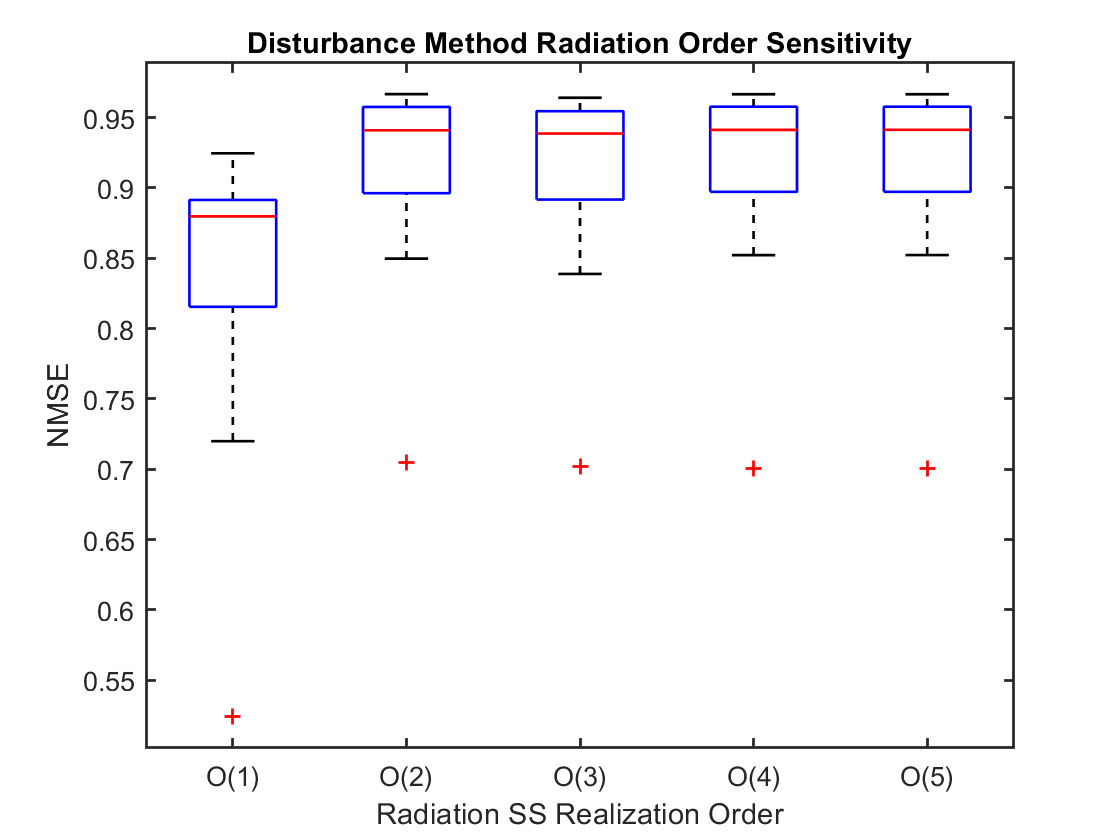}
    \end{center}
    \caption{The excitation force estimation NMSE using the disturbance estimation method of each run shown for various orders of the state space radiation realization. O(*) refers to the order of the state space realization of the radiation force. Outliers are denoted with a red $+$.}
    \label{NMSE_DistRad}
\end{figure}

It is shown in both Figs. \ref{NMSE_DirRad} and \ref{NMSE_DistRad} that using a first order realization for the radiation results in a significant decrease in performance. Note that for the second through fifth order realizations the performance of the estimator remains fairly constant. However, as mentioned above, it was shown in \cite{Davis2018} that the fourth order radiation force is the smallest realization that is viable for identifying a missing drag force for this heaving float. As a result, the fourth order radiation force is used throughout the rest of the simulations in this work.

\subsection{Sensitivity to Measurement Noise}
\begin{figure}[h!]
    \begin{center}
       \includegraphics[width= 85mm]{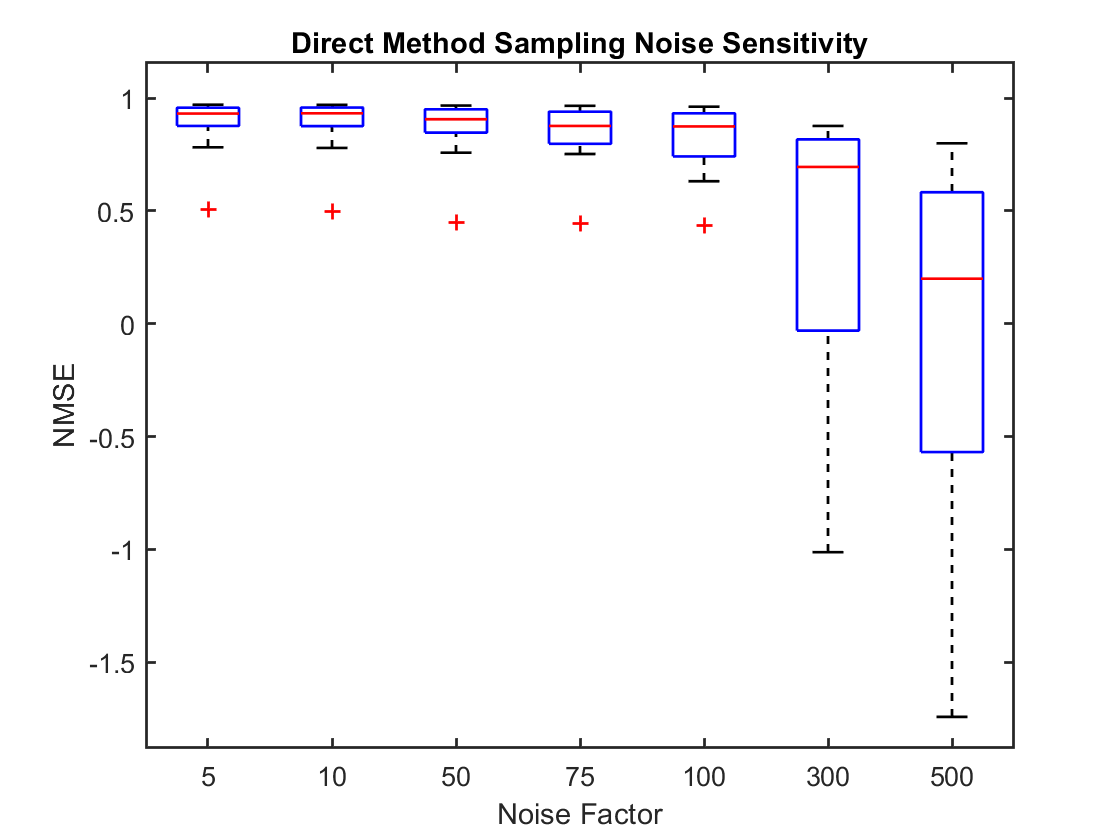}
    \end{center}
    \caption{The excitation force estimation NMSE using the direct estimation method of each run shown for various added measurement noises. Outliers are denoted with a red $+$.}
    \label{NMSE_noiseDirect}
\end{figure}

\begin{figure}[h!]
    \begin{center}
       \includegraphics[width= 85mm]{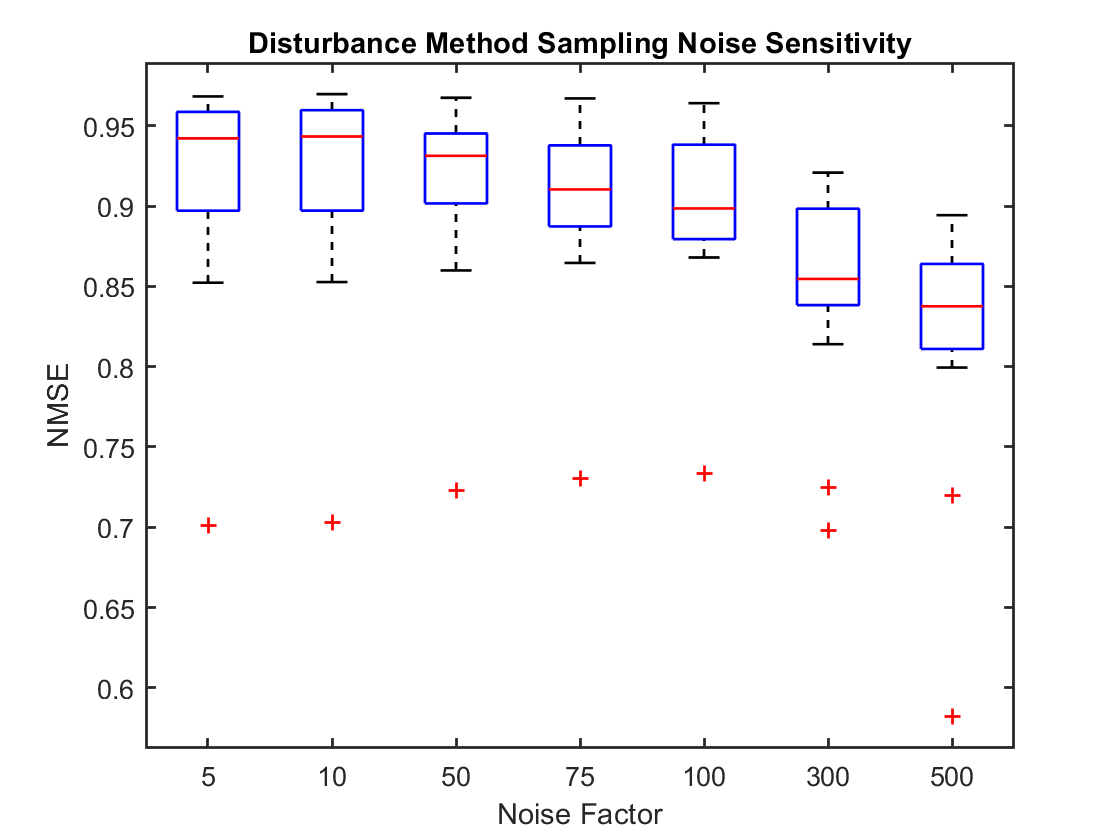}
    \end{center}
    \caption{The excitation force estimation NMSE using the disturbance estimation method of each run shown for various added measurement noises. Outliers are denoted with a red $+$.}
    \label{NMSE_noiseDisturb}
\end{figure}

The experiments described in \cite{Bosma2015} obtained high quality position measurements by using a Qualysis data acquisition system. The measurement noise used in the EKFs is very small, calculated from the residuals of the position measurements. The standard deviation of the measurement noise for position and velocity was $v=6.075\times10^{-4}$. To investigate the sensitivity of the estimation algorithms to measurement noise an artificial, zero mean, Gaussian noise term was added to the measurements. Figures \ref{NMSE_noiseDirect} and \ref{NMSE_noiseDisturb} show the NMSE for the direct estimation and disturbance estimation methods, respectively, as artificial measurement noise is added to the experimental measurements. The actual measurement noise is very small, the artificial measurement noise, $v_a$ is described by a noise factor that is an integer multiple of the experimental measurement noise. Both Figs. \ref{NMSE_noiseDirect} and \ref{NMSE_noiseDisturb} show the resilience of the EKF, as no appreciable performance degradation is shown until the artificial noise is several orders of magnitude larger than the measurement noise.

\section{Conclusion} \label{sec:conclusion}
This paper presents the implementation of an Extended Kalman Filter (EKF) to estimate the wave excitation force of a Wave Energy Converter (WEC). Many advanced control strategies require the wave excitation force that is acting on the WEC to determine the control input. However, real time measurement of the wave excitation force is typically not viable for ocean deployments. To provide information about the wave excitation force, Kalman filters are implemented as optimal state observers which are configured to estimate the wave excitation force along with the dynamics of the WEC. Prior to this work, typical implementations of the Kalman filter have been restricted to linear Kalman filters and have been tested with simulations rather than experimental data.

Two different estimation methods, making use of the nonlinear Kalman filter, are described in this work, both of which are tested with experimental data from a 1:10 scale wave tank test of a floating power system which closely resembles a point absorbing float. The \textit{direct estimation method} relies on modeling the excitation force as a single state with no dynamics, allowing for the ``random walk'' of the Kalman filter to determine a statistically significant estimate of the excitation force. The \textit{disturbance estimation method} models the wave excitation force as a sum of sine waves with the appropriate derivatives modeled as states. An equal energy method is used to determine the amplitudes and frequencies of the harmonic components. The quantity of the harmonic excitation terms is varied and discussed in this work.

The potential based time domain integro-differential equation of motion, used in this work, is a widely applied lumped parameter model of WEC bodies. This model includes a Morison drag term which is nonlinear and requires knowledge of the incident wave velocity to describe the drag acting on the object. The nonlinearity of this model requires the use of the EKF rather than the simpler linear Kalman filter. Much like the wave excitation force, the incident wave velocity is typically unknown in WEC deployments. A simple approximation of the incident wave velocity is developed and tested in this work, which enables the full implementation of the nonlinear equation of motion. The normalized mean square error between the estimated excitation force and true excitation force is evaluated using seventeen different wave climates. The sensitivity of the excitation force estimation error is evaluated for variation in the model of radiation force, measurement noise, sampling rate as well as wave climate. 

Both of the estimation methods discussed in this work were shown to provide good estimates of the excitation force using experimental data that would be available during a WEC deployment. The harmonic oscillator model of the wave excitation force utilized in the disturbance estimation method is a more realistic model of the force, which results in some improvements in estimation accuracy. However, the disturbance model adds greatly to the computational complexity of the extended Kalman filter while providing modest improvements. The excitation force estimation of both methods is shown to be quite resilient to a wide range of situations. This work shows that a common, nonlinear, lumped parameter model, previously unused for wave excitation force estimation, can be used to accurately estimate experimental excitation forces through the use of an extended Kalman filter.

\clearpage
\bibliography{mybibfile}

\end{document}